\documentclass[10pt]{article}
\usepackage{graphicx}
\begin{document}
\title{Polarization of cosmic microwave background
scattered by moving flat protoobjects}
\author{N.N.~Shakhvorostova$^1$, S.I.~Grachev$^2$, V.K.~Dubrovich$^3$}
\maketitle

\begin{center}
$^1$Astro-Space Center, Lebedev's Physical Institute, Moscow, Russia\\
$^2$Saint-Petersburg State University, Sobolev Astronomical Institute,
Saint-Petersburg, Russia\\
$^3$Special Astrophysical Observatory, Saint-Petersburg Department,
Saint-Petersburg, Russia\\
\end{center}

\vskip 10 pt

The work is devoted to the investigations of possible observational
manifestations of
protoobjects related to ``dark ages" epoch $(10<z<1000)$, before formation of
self-luminous galaxies and stars. These objects can distort the cosmic
microwave background. Formation of these objects is described in the ``pancake
theory" and in the model of ``hierarchic clustering". According to these 
theories we may consider these protoobjects as flat layers.

We consider both Thomson (with Rayleigh phase matrix) and resonance (for
complete frequency redistribution) scattering of cosmic microwave background 
radiation by moving flat layer. The appearing anisotropy and polarization of 
cosmic microwave radiation are calculated for a wide range of a layer optical
thicknesses (from optically thin layer to optically thick one). Analytical 
solutions
 are also obtained for the case of an optically thin layer and are 
compared
 with the numerical ones.

\section{Introduction}

We consider a special period in the Universe evolution extending from 
recombination epoch $(z=1000)$ till the time of self-emitting
stars and galaxies formation $(z=10)$. This period is called ``dark ages" and
the question about possible observational manifestations of dark objects from
that period is actual. One of the most effective ways to solve this problem is
to try to observe the cosmic microwave background (CMB) distortions caused
by these objects, as it was firstly reported in \cite{1}. The character of such
distortions is defined by the physical properties of ``dark ages" objects, 
which are naturally the protoclusters of galaxies \cite{2}.

Now two basic theories of galaxy clusters formation are being discussed. The
first one, so called ``pancake theory", was suggested by Ya.B. Zeldovich (see 
\cite{3} for example) and further developed in some papers (see \cite{4}, 
\cite{5}).
 According to this
theory, the substance is collapsing into flat formations --- ``pancakes"
(protoclusters). Initial shock waves forming in this process can be modelled
as flat layers. According to the second theory, ``the model of hierarchic
clustering" \cite{6}, small structures are merging into larger ones (``the 
boxes"). These structures contain a large number of small dense objects merging 
to form present galaxies and diffuse gas substance. The difference between 
these two galaxy clusters formation models consists only in the value of mass 
ratio of dense objects and diffuse substance. In the first theory gas 
``pancakes" contain the most part of the total mass, and in the second one the
most but not the whole mass is contained in dense objects. Therefore, the gas 
mass is not equal to zero and its spatial distribution as flat ``pancakes" 
takes place in both theories and we may consider them as flat layers.

Possible distortions of CMB appear due to the fact that these flat layers can
move relatively CMB with peculiar velocities (see \cite{7} for example). 
Due to the Doppler effect caused by this motion the CMB becomes anisotropic
in the reference frame of the layer. This anisotropy, which is axially
symmetric with respect to the direction of the motion, leads to polarization 
and intensity changes of scattered radiation. These distortions belong to 
the ``secondary CMB distortions". Their differences from ones forming at 
the hydrogen recombination time $(z=1000)$ consist in other angular scales and 
possibly in a more complicated spectrum. In particular, ``secondary 
fluctuations" can appear not only due to the scattering on electrons, but on
primary molecules as well \cite{8}, \cite{9}, \cite{10}. 

Peculiar velocities of substance large scale fluctuations for a wide
range of cosmological models can be estimated by the formula \cite{11}, 
\cite{12}:
\begin{equation}
\label{1.1}
v_p=600\,(1+z)^{-1/2}\,\,{\rm km/s}\mbox{,}
\end{equation}
where the numerical coefficient is obtained as a result of a large number
of velocity estimations of nearby clusters of galaxies. The velocity dependence
on $z$ is defined by the general law of the massive body motion in
the expanding Universe. The calculation of the optical depth is different for
scattering on free electrons (Thomson scattering) and for resonance scattering
on atoms, ions or molecules. Thomson optical depth does not depend on the
wavelength and is defined by the size of density fluctuation and electron
density. The last quantity sufficiently depends on the redshift: after the
hydrogen recombination at $z=1300$ the fractional electrons density 
falls from the unity to the value about $0.01\%$ at $z=100$. But after that
time the number of free electrons rises due to the secondary ionization the
details of which are being investigated by several authors and may
sufficiently differ in various models (see \cite{13} for example).
Observational restrictions based on the power spectrum of primary CMB 
fluctuations give for the average optical thickness of such objects
$\tau_0<0.1$. However, it can achieve unity for particular objects.

Existence of free electrons and protons leads automatically to the appearance
of bremsstrahlung not related with the peculiar velocity. The role of such
emission in the secondary ionization epoch was investigated in \cite{14} where
it was pointed out that there is a quite wide range of fluctuation parameters 
for 
which the role of this process is very small.

As concerned a resonance scattering in molecular lines there is a strong 
dependency of observed effect on the frequency. The estimation of the optical
thickness of the layer (averaged over the line) $\tau_0$ in this case is much 
more complicated problem since the density of different molecules strongly 
depends on physical processes in a given epoch \cite{9}. In principle, the 
estimation of second type shocks efficiency (excitation and disactivation by 
collisions with electrons and atoms) is necessary here. However, we may 
neglect such processes for 
the considered epoch because of a very low absolute concentration of colliding
particles (numerical estimates are made in \cite{9}). Thus, in a first 
approximation we may consider only the Doppler mechanism of CMB distortions 
formation. CMB interaction with molecules having high dipole moments will be 
most effective. From this point of view the main attention was paid to the 
molecules $HD^+$, $HeH^+$, $LiH$ and some other ones \cite{1}, \cite{15}.
A large number of $HeH^+$ molecules can be formed in the shock wave
at the stage of nonlinear collapse of primary ``pancakes". This way
to observe the large scale distribution of substance at $z\sim 10$
appears to be possible. The role of other molecules is many times
less (see \cite{10} for example). The lines of rotational and vibrational
transitions of such molecules at $z\sim 200$ can be situated in a
millimeter range. The discovery of two lines of rotational-vibrational
array should allow to define the type of molecule and $z$.

The effects of CMB scattering on free electrons in 
the hot intergalactic gas of rich clusters of galaxies were considered in the 
papers by Zeldovich and Sunyaev \cite{17}, \cite{18}, \cite{19}, where the 
analytical estimates of CMB polarization caused both by the gas motion and 
by the intrinsic anisotropy of CMB were obtained. Those papers were principally 
based on a single scattering assumption which, however, leads to polarization 
which is proportional only to $(v/c)^2$. Analytical estimate of $O(v/c)$ effects
with account of double scattering for moving spherical objects was made in 
\cite{17}. 

The primary goal of our work is the numerical calculation of CMB distortion
effects depending on the layer parameters with no restriction concerning to 
the number of scatterings taken into account.
We consider both Thomson (with Rayleigh phase matrix) and resonance (for
complete frequency redistribution) scattering of cosmic mircowave background by
moving flat layer. The appearing anisotropy and polarization of cosmic 
microwave radiation are calculated for a wide range of a layer opticall 
thicknesses (from optically thin layer to optically thick one). Analytical 
solutions
 are also obtained for the case of an optically thin layer and are 
compared
 with the numerical solutions. Since the electron number density 
after the recombination epoch becomes too small, this mechanism will be more 
effective at the beginning of nonlinear stage of protoobjects evolution at the
``secondary ionization" epoch $(10<z<50)$ (see \cite{11} for example). The 
molecular scattering can play sufficient role at $z<150$.

The predicted effects can be observed in a wide range of angular sizes
which depend on cosmological model parameters and may vary from dozens
of angular seconds to dozens of angular minutes.

\section{Rayleigh scattering}

Let us consider plane-parallel horizontally homogeneous layer moving as a whole
with a velocity {\bf v} (with regard to CMB) at the direction of an outer 
normal {\bf n} to the layer. According to eq. (\ref{1.1}) peculiar velocities
of such objects
are about 150 km/s. Thus, we may assume $v/c\ll 1$ and disregard the terms
of the order of $(v/c)^2$ in further calculations. The layer is illuminated 
by CMB with the intensity (in its own reference frame) described by Planck 
function with the temperature $T=2.7K$. Having transformed to the
layer reference frame the CMB intensity is given by
\begin{equation}
\label{2.1}
I(\mu,\nu)=\frac{2h\nu^3}{c^2}\frac{1}{e^{\gamma(1+\mu\beta)h\nu/kT}-1}\mbox{,}
\end{equation}
where $\nu$ is the radiation frequency, $\mu$ is the cosine of the angle
between the velocity and radiation propagation directions in the layer 
reference frame, $\beta=v/c$, $\gamma=1/\sqrt{1-\beta^2}$.

Since we have azimuthal-symmetric picture, the field of radiation is
completely described by two-component Stokes vector ${\bf i}=(I,Q)^{\rm T}$,
where ``{\rm T}" means matrix transposition, $I=I(r,\mu,\nu)$,
$Q=Q(r,\mu,\nu)$, $r$ is the distance from one of the layer borders.

In the case of Rayleigh scattering we will use the optical
depth $\tau$ as the geometric variable instead of $r$: $d\tau =-n_e\sigma dr$
and denote the layer optical thickness as $\tau_0$. Here $n_e$ is the electron
density, $\sigma$ is the Thomson cross-section. Since the picture is
independent of frequency, the Stokes vector can be written as
${\bf i}(\tau,\mu)=(I(\tau,\mu),Q(\tau,\mu))^{\rm T}$. This vector is the 
solution of radiative transfer equation (\cite{20}, \cite{21})
\begin{equation}
\label{maineq}
\mu\frac{\partial{\bf i}(\tau,\mu)}{\partial\tau}={\bf i}(\tau,\mu)-
(1/2)\int_{-1}^1\hat{P}(\mu,\mu'){\bf i}(\tau,\mu')d\mu'-
{\bf s}^*(\tau,\mu)\equiv {\bf i}(\tau,\mu)-{\bf s}(\tau,\mu)
\end{equation}
with the boundary conditions
\begin{equation}
{\bf i}(0,\mu)={\bf i}_1(\mu),\quad\mu<0;\qquad
{\bf i}(\tau_0,\mu)={\bf i}_2(\mu),\quad\mu>0,
\end{equation}
where ${\bf i}_1(\mu)$ and ${\bf i}_2(\mu)$ are the Stokes vectors of 
radiation illuminating the layer from outside at the boundaries $\tau=0$ and 
$\tau=\tau_0$, respectively. Here ${\bf s}^*$ characterizes primary sources 
distribution in the layer and the phase matrix $\hat{P}(\mu,\mu')$ is, in
general, the 
superposition of the Rayleigh phase matrix $\hat{P}_{\rm R}$ and the phase
matrix of isotropic scattering $\hat{P}_{\rm I}$: $\hat{P}=(1-W)\hat{P}_{\rm I}
+W\hat{P}_{\rm R}$. Here $W$ is the depolarization parameter which is usually
in the range [0,1]. For the phase matrix the following factorization is 
obtained (see \cite{21}):
$\hat{P}(\mu,\mu')=\hat{A}(\mu)\hat{A}^{\rm T}(\mu')$, where the matrix
\begin{equation}
\label{2.4}
\hat{A}(\mu)=\left(\begin{array}{cc}
                   1& b(1-3\mu^2)\\
                   0& 3b(1-\mu^2)\\
                   \end{array}\right).
\end{equation}
Here and below $b=\sqrt{W/8}$. In the case of Rayleigh scattering considered 
here the depolarization parameter $W$ is equal to 1.

Taking into account a single scattering of an outer radiation and using the 
phase matrix factorization, we reduce the problem to the solution of eq.
(\ref{maineq}) with the free term
\begin{equation}
\label{2.5}
{\bf s}^*(\tau,\mu)=(1/2)\hat{A}(\mu)\int_0^1\hat{A}^{\rm T}(\mu')
\left[e^{-(\tau_0-\tau)/\mu'}{\bf i}_2(\mu')+e^{-\tau/\mu'}{\bf i}_1(-\mu')
\right]d\mu'
\end{equation}
and zero boundary conditions
\begin{equation}
\label{2.6}
{\bf i}(0,\mu)=0,\quad\mu<0;\qquad {\bf i}(\tau_0,\mu)=0,\quad
\mu>0.
\end{equation}
For ${\bf i}_1(\mu)$ and ${\bf i}_2(\mu)$ we use the expansion of
background radiation in the layer reference frame (\ref{2.1}) with the 
accuracy of the order of $O(v/c)$:
\begin{equation}
\label{2.7}
{\bf i}_1(\mu)={\bf i}_2(\mu)\sim B(\nu,T)\left[1-(v/c)a_\nu\mu\right]
{\bf e}_1,
\end{equation}
where $a_\nu=xe^x/(e^x-1)$, $x=h\nu/kT$, ${\bf e}_1=(1,0)^{\rm T}$.

According to eq. (\ref{2.7}) we devide the both sides of the main 
equation (\ref{maineq}) by the depth independent factor $B(\nu,T)$ so that
below the Stokes vector ${\bf i}(\tau,\mu)$ and the vector source function
${\bf s}(\tau,\mu)$ are dimensionless (measured in the units of $B(\nu,T)$).

The factorization of the phase matrix $\hat{P}(\mu,\mu')$ leads evidently to
factorization of the source term in the righthandside of eq. (\ref{maineq}):
${\bf s}(\tau,\mu)=\hat{A}(\mu){\bf s}(\tau)$. Then we use the procedure 
described in \cite{21} to reduce the problem to the solution of the linear  
integral equation for a vector source function ${\bf s}(\tau)$ depending
only on the optical depth. Finally we obtain the following expression for the
Stokes vector of radiation emerging through the boundary 
$\tau=0$ (for the observer reference frame):
\begin{equation}
\label{2.8}
{\bf i}(0,\mu)={\bf e}_1+(v/c)a_\nu\mu\left(1-e^{-\tau_0/\mu}
\right){\bf e}_1+(v/c)a_\nu\hat{A}(\mu)\int_0^{\tau_0}e^{-\tau/\mu}
{\bf s}(\tau)d\tau/\mu,\quad \mu>0,
\end{equation}
where ${\bf s}(\tau)=(s_{\rm I}(\tau),s_{\rm Q}(\tau))^{\rm T}$ satisfies to
the vector integral equation
\begin{equation}
\label{2.9}
{\bf s}(\tau)=(1/2)\int_0^{\tau_0}\hat{K}(|\tau-\tau'|){\bf s}(\tau')
d\tau'+{\bf s}^*(\tau)
\end{equation}
with the core matrix (see \cite{21}
)
\begin{equation}
\label{2.10}
\hat{K}(\tau)=\left(\begin{array}{cc}
                  E_1(\tau)& b[E_1(\tau)-3E_3(\tau)]\\
  b[E_1(\tau)-3E_3(\tau)]& 2b^2[5E_1(\tau)-12E_3(\tau)+9E_5(\tau)]\\
  \end{array}\right)
\end{equation}
and the primary source term
\begin{equation}
\label{2.11}
{\bf s}^*(\tau)=(1/2)\{E_3(\tau)-E_3(\tau_0-\tau),b[E_3(\tau)-
E_3(\tau_0-\tau)-3(E_5(\tau)-E_5(\tau_0-\tau))]\}^{\rm T}.
\end{equation}
Here $b=\sqrt{1/8}$, ${\displaystyle E_n(\tau)=\int_0^1e^{-\tau/\mu}
\mu^{n-2}d\mu}$ is the $n$-th integral exponent. 

\begin{figure}[t]
\vspace*{-6cm}

\centering
\resizebox{1.0\textwidth}{!}{\includegraphics{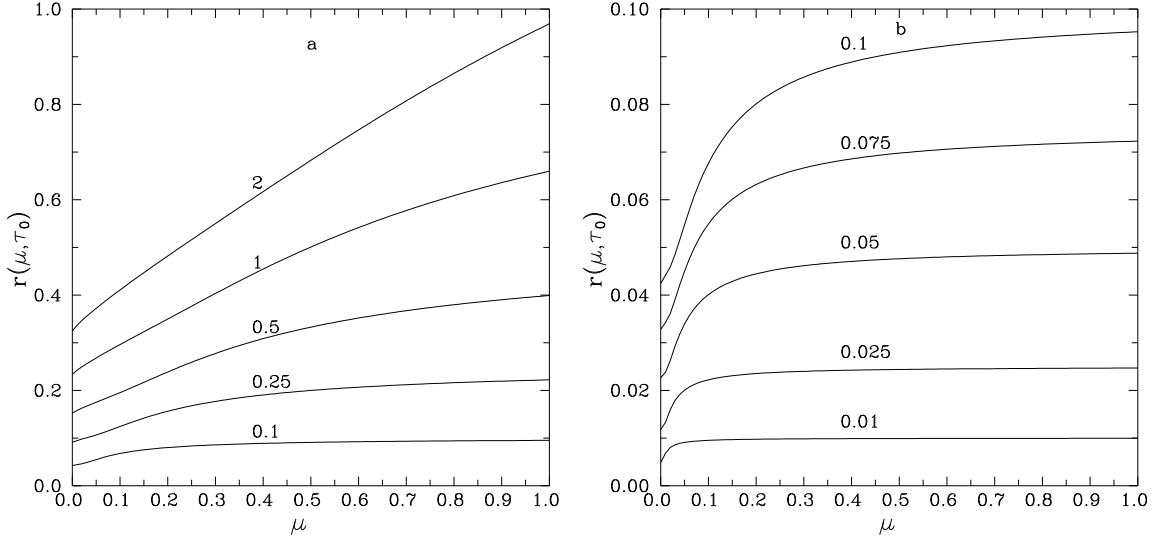}}

\vspace{-11cm}

\caption{Profiles of intensity change for Rayleigh scattering for different 
values of $\tau_0$ marked near the curves.}
\end{figure}

For the intensity change $\Delta I/I_0\equiv (I-B)/B$ and for the polarization
degree of radiation emerging through the boundary $\tau=0$ we obtain 
from eq. (\ref{2.8})
\begin{equation}
\label{2.12}
\begin{array}{cc} 
\Delta I/I_0=(v/c)a_{\nu}r(\mu,\tau_0)+{\rm O}((v/c)^2),\\
  Q/I=-(v/c)a_{\nu}P(\mu,\tau_0)+{\rm O}((v/c)^2),\\
\end{array}
\end{equation}
where the profiles of intensity change and polarization are
\begin{equation}
\label{2.13}
\displaystyle
\begin{array}{cc}
{\displaystyle
r(\mu,\tau_0)=\mu\left(1-e^{-\tau_0/\mu}\right)+\int_0^{\tau_0}e^{-\tau/\mu}
\left[s_{\rm I}(\tau)+(1/\sqrt{8})(1-3\mu^2)s_{\rm Q}(\tau)\right]d\tau/\mu,}\\
{\displaystyle
P(\mu,\tau_0)=-(3/\sqrt{8})(1-\mu^2)\int_0^{\tau_0}e^{-\tau/\mu}
s_{\rm Q}(\tau)d\tau/\mu.}\\
\end{array}
\end{equation}
For the maximum polarization degree which is reached at $\mu=0$ and for the
corresponding intensity change we have from eq. (\ref{2.13})
\begin{equation}
\label{2.14}
\begin{array}{cc}
P(0,\tau_0)=-(3/\sqrt{8})s_{\rm Q}(0),\\
r(0,\tau_0)=s_{\rm I}(0)+(1/\sqrt{8})s_{\rm Q}(0).\\
\end{array}
\end{equation}

For the radiation emerging through the boundary $\tau=\tau_0$ we obviously
should change the sign of $v$ in the formulae above i.e. to change 
the sign of polarization and intensity variation.

Integral equation (\ref{2.9}) was solved numerically by discretization
at some $\tau$ grid and reducing to the system of linear algebraic equations
for the values of the source function at the knots of the grid. These values
are used then to calculate the profiles of the intensity change and
polarization through eq. (\ref{2.13}). The results are shown 
in Figs. 1 and 2.

\begin{figure}[t]
\vspace*{-6cm}

\centering
\resizebox{1.0\textwidth}{!}{\includegraphics{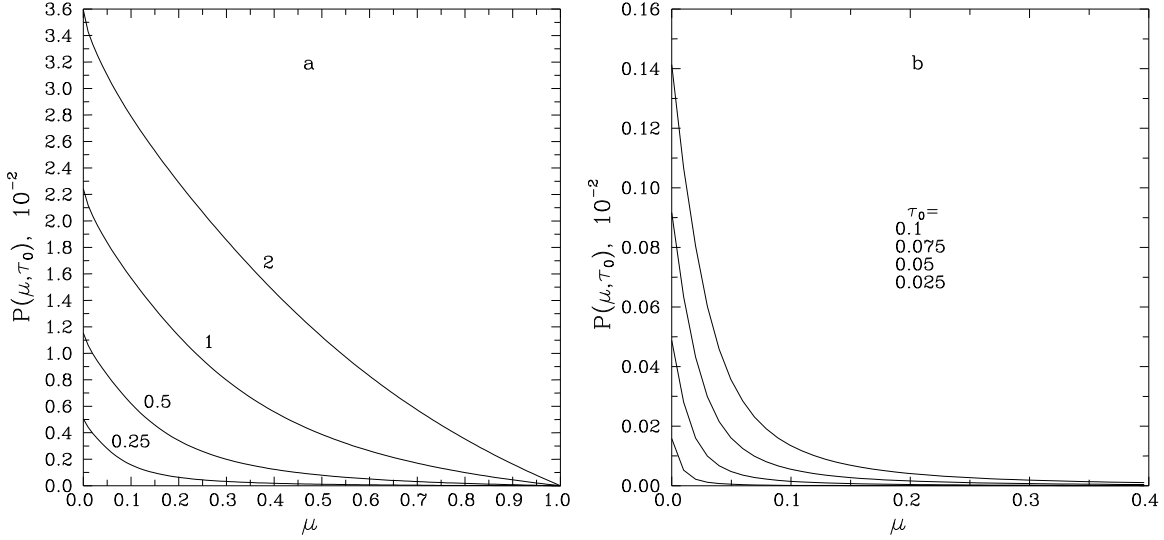}}

\vspace{-11cm}

\caption{Profiles of polarization for 
Rayleigh scattering for different values of $\tau_0$ marked near the curves 
(a)
 or listed in the same order as the curves follow, from up to down, (b).
}
\end{figure}

As we can see in Fig. 2b there is a strong dependency of polarization degree 
on $\mu$ for small $\tau_0$. For $\tau_0\ll 1$ we obtain the following
expansions for $s_{\rm I}(\tau)$ and $s_{\rm Q}(\tau)$ with the accuracies
up to the first and second order on the optical depth respectively:
\begin{equation}
\label{2.15}
{\bf s}(\tau)=\left(\begin{array}{cc}
s_{\rm I}(\tau) \\
s_{\rm Q}(\tau) \\
\end{array}\right)\sim\left(\begin{array}{cc}
-\tau+\tau_0/2 \\
(\tau_0/8\sqrt{2})[(\tau_0-2\tau)(C+1/2)-
\tau\ln\tau+(\tau_0-\tau)\ln(\tau_0-\tau)] \\
\end{array}\right)\mbox{,}
\end{equation}
where $C=0.577216$ is the Euler constant. Substitution of these expansions
into eqs. (\ref{2.13}) and (\ref{2.14}) gives us the analytical
expressions of the intensity change and polarization profiles
for $\tau_0\ll 1$:
\begin{equation}
\label{2.16}
r(\mu,\tau_0)\sim\left\{\begin{array}{ll}
 \tau_0/2,\quad \mu\ll\tau_0,\\
 \tau_0,\quad \mu\gg\tau_0,\\
\end{array}\right.
\end{equation}

\begin{equation}
\label{2.17}
P(\mu,\tau_0)\sim\left\{\begin{array}{ll}
P(0,\tau_0)\sim-(3/32)\tau_0^2(\ln\tau_0+C+1/2),\quad\mu\ll\tau_0,\\
-(1/64)\tau_0^4(1-\mu^2)\mu^{-2}(\ln\tau_0+C+2/3),\quad\mu\gg\tau_0.\\
 \end{array}\right.\mbox{.}
\end{equation}

Eq. (\ref{2.17}) explains the strong dependency of polarization degree on
$\mu$. It is confirmed by numerical calculations. For example,
at $\tau_0=0.01$ its error is less than $4\%$ for $\mu>0.09$, at $\tau_0=0.025$
is less than $4\%$ for $\mu>0.14$ and at $\tau_0=0.1$ is less than $10\%$ for
$\mu>0.13$. 
We can obtain even more precise approximation taking into account the
terms of larger order with respect to $\tau_0$. Thus, for $P(0,\tau_0)$ we have
\begin{equation}
\label{2.18}
P(0,\tau_0)\approx -(3/32)\tau_0^2[(1+1.25\tau_0)\ln\tau_0+C+1/2]\mbox{,}
\end{equation}
the error of which is less than $1\%$ for $\tau_0<0.1$.

An approximate analytical estimate of the maximum polarization degree for a
homogeneous sphere with a small optical radius $\tau_0$ was obtained 
by Sunyaev and Zel'dovich \cite{17}:
$(Q/I)_{\rm max}=\pm (v_t/c)a_{\nu}\tau_0^2/10$, where $v_t$ is the transversal
velocity. However, the accuracy of this estimate is unknown because it was
not compared with numerical calculations.

Obtained dependencies of CMB intensity and polarization on layer parameters
allow us to model observational manifestations of these objects. In the
protoobject directions we would registrate increase or decrease of CMB
intensity depending on the sign of velocity $v$. Having measured the
intensity change, we may estimate the $(v/c)\tau_0$ value through the eqs.
(\ref{2.12}) and (\ref{2.16}). Numerical analysis of polarization in couple 
with intensity
 variation data would permit to obtain the values of velocity
and optical thickness of the layer separately.

\section{Resonance scattering}

Let us consider now the case of CMBR scattering in a spectral line, say, in
a line of $HeH^+$ with the laboratory wavelength $149 \mu$. At $z=150$ this 
corresponds to the wavelength 2.25 cm.

We use the model of two-level atoms with the transition frequency
$\nu_{12}$. Let $x$ be the dimensionless frequency:
$x=(\nu-\nu_{12})/\Delta\nu_D$, where $\Delta\nu_D$ is the Doppler width
of the line. Let $\phi(x)$ be the absorption coefficient profile normalized
as follows: $\int_{-\infty}^{\infty}\phi(x)dx=1$.

Since we have axial symmetry again, the Stokes vector {\bf i} will
have only two components ($I$ and $Q$). But now it will depend not only on
coordinate and direction ($\tau$ and $\mu$) but on frequency $x$. The
radiative transfer equation is written as (see \cite{22} for example):
\begin{eqnarray}
\label{3.1}
&{\displaystyle \mu\frac{\partial{\bf i}(\tau,\mu,x)}{\partial\tau}=
\phi(x){\bf i}(\tau,\mu,x)-
\frac{\lambda}{2}\int_{-1}^{1}d\mu'\int_{-\infty}^{\infty}\hat{R}(\mu,x;\mu',
x'){\bf i}(\tau,\mu',x')dx'-{\bf s}^*(\tau,\mu,x)\equiv}& \nonumber\\ 
&\equiv\phi(x){\bf i}(\tau,\mu,x)-{\bf s}(\tau,\mu,x).&
\end{eqnarray}

The matrix $\hat{R}(\mu,x;\mu',x')$ describes the redistribution over
frequencies, angles and polarization conditions at a single scattering, 
$\lambda$ is the single scattering albedo, ${\bf s}^*(\tau,\mu,x)$ is the 
primary source function vector. If there is no primary sources embeded in the 
layer it is defined by illumination from outside:
\begin{equation}
\label{3.2}
{\bf s}^*(\tau,\mu,x)=\frac{\lambda}{2}\int_{-\infty}^{\infty}dx'\int_{0}^{1}
\hat{R}(\mu,x;\mu',x')\left[e^{-(\tau_0-\tau)\phi(x')/\mu'}{\bf i_2}
(\mu',x')+
e^{-\tau\phi(x')/\mu'}{\bf i_1}(-\mu',x')\right]d\mu'\mbox{,}
\end{equation}
where ${\bf i_1}(\mu,x)={\bf i_2}(\mu,x)\sim B(\nu_{12},T)[1-(v/c)a_{\nu_{12}}
\mu]{\bf e}_1$, $a_{\nu_{12}}=x_{12}/(1-e^{-x_{12}})$, $x_{12}=h\nu_{12}/kT$,
${\bf e}_1=(1,0)^{\rm T}$. As in the case of Rayleigh scattering we use below
dimensionless Stokes vector and vector source function (measured in the
units of $B(\nu_{12},T)$).

We use the assumption of complete frequency redistribution, according to
which (see \cite{22})
$\hat{R}(\mu,x;\mu',x')=\phi(x)\phi(x')\hat{P}(\mu,\mu')=\phi(x)\phi(x')\hat{A}
(\mu)\hat{A}^{\rm T}(\mu')$,
where $\hat{A}(\mu)$ is defined by eq. (\ref{2.4}). The value of 
depolarization parameter $W$ in that equation is defined by the quantum numbers 
of the total angular momentum of the upper and the lower levels of transition 
(see \cite{20} for example). Polarization becomes smaller as $W$ decreases.
$W=1$ corresponds to dipole scattering. In this case the phase matrix is
the same as for the Rayleigh scattering.

The factorization of the redistribution matrix $\hat{R}(\mu,x;\mu',x')$ leads 
to factorization of the source term in the righthandside of eq. (\ref{3.1}):
${\bf s}(\tau,\mu,x)=\phi(x)\hat{A}(\mu){\bf s}(\tau)$. Then we use the 
procedure described in \cite{22} to reduce the problem to the solution of the
linear integral equation of the type (\ref{2.9}) for a vector source 
function ${\bf s}(\tau)$ depending only on the optical depth. But now 
the elements of the core matrix $\hat{K}(\tau)$ in eq. (\ref{2.9}) are 
(see \cite{22}):
\begin{equation}
\label{3.6}
\begin{array}{lll}
{\displaystyle
K_{11}(\tau)=\lambda\int_{-\infty}^{\infty}\phi^2(x)E_1(\phi(x)\tau)dx,}\\
{\displaystyle
K_{12}(\tau)=K_{21}(\tau)=b\lambda\int_{-\infty}^{\infty}\phi^2(x)\left[E_1
(\phi(x)\tau)-3E_3(\phi(x)\tau)\right]dx,}\\
{\displaystyle
K_{22}(\tau)=2b^2\lambda\int_{-\infty}^{\infty}\phi^2(x)\left[5E_1(\phi(x)\tau)
-12E_3(\phi(x)
\tau)+9E_5(\phi(x)\tau)\right]dx\mbox{.}}\\
\end{array}
\end{equation}

\begin{figure}[t]
\vspace*{-6cm}

\centering
\resizebox{1.0\textwidth}{!}{\includegraphics{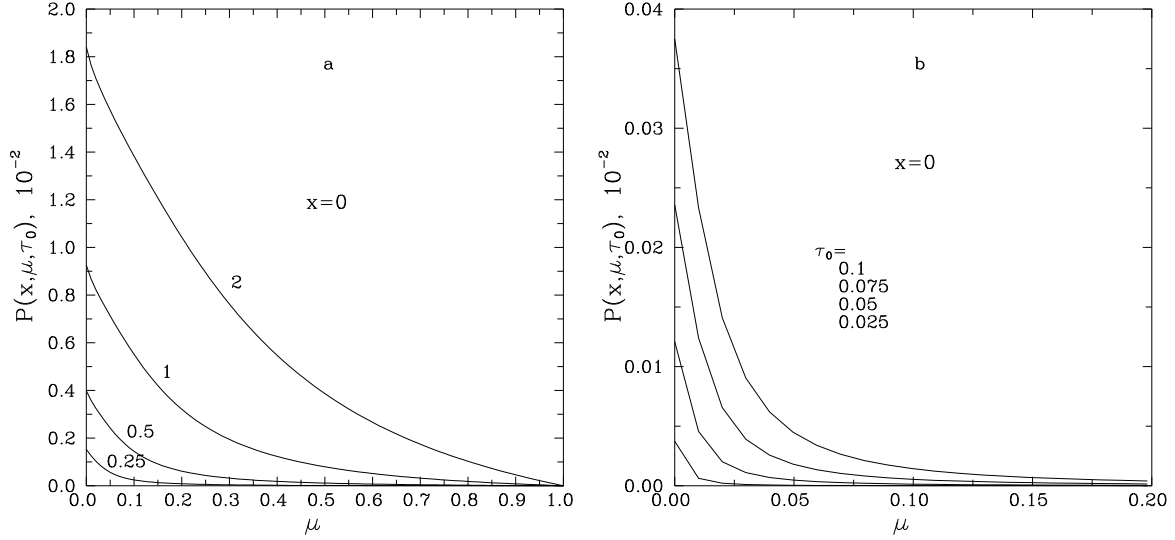}}

\vspace{-11cm}

\caption{Angular profiles of polarization in a 
center of a spectral line for different values of $\tau_0$ marked near the
curves (a) or listed in the same order as the curves follow, 
from up to down, 
(b).}
\end{figure}

Finally we obtain the following expression for the Stokes vector of emerging 
radiation:
\begin{equation}
\label{3.3}
\begin{array}{cc}
{\displaystyle
{\bf i}(0,\mu,x)={\bf e}_1+(v/c)a_{\nu_{12}}\mu
\left(1-e^{-\tau_0/\zeta}\right){\bf e}_1-(1-\lambda)[1+(v/c)
a_{\nu_{12}}\mu]
\hat{A}(\mu)\int_{0}^{\tau_0}e^{-\tau/\zeta}{\bf s}_0(\tau)d\tau/\zeta+}\\
{\displaystyle
+(v/c)a_{\nu_{12}}\hat{A}(\mu)\int_{0}^{\tau_0}e^{-\tau/\zeta}
{\bf s}(\tau)d\tau/\zeta,\quad \mu>0,}\\
\end{array}
\end{equation}
where $\zeta=\mu/\phi(x)$, ${\bf s}_0(\tau)$ is the solution of type 
(\ref{2.9}) equation with the free term ${\bf s}^*={\bf e}_1$, and 
${\bf s}(\tau)= (s_{\rm I}(\tau),s_{\rm Q}(\tau))^{\rm T}$ is
the solution of the same equation but with the free term ${\bf s}^*=
(s_{\rm I}^{*},s_{\rm Q}^{*})^{\rm T}$: 
\begin{equation}
\label{3.4}
s_{\rm I}^{*}=\frac{\lambda}{2}\int_{-\infty}^{\infty}\phi(x)\left[E_3(\phi(x)
\tau)-E_3(\phi(x)(\tau_0-\tau))\right]dx\mbox{,}
\end{equation}
\begin{equation}
\label{3.5}
s_{\rm Q}^{*}=\frac{\lambda}{2}b\int_{-\infty}^{\infty}\phi(x)\left\{\left[E_3(
\phi(x)\tau)-E_3(\phi(x)(\tau_0-\tau))\right]-
3\left[E_5(\phi(x)\tau)-E_5(\phi(x)(\tau_0-\tau))\right]\right\}dx\mbox{.}
\end{equation}

Since for the most part of molecules the value of $1-\lambda$ is about 
$10^{-9}$ and the value of
$v/c$ is about $10^{-4}$, we may neglect the term proportional to 
$1-\lambda$ in eq. (\ref{3.3}).
However, the account of this term does not bring any difficulties.

\begin{figure}[t]
\vspace*{-6cm}

\centering
\resizebox{1.0\textwidth}{!}{\includegraphics{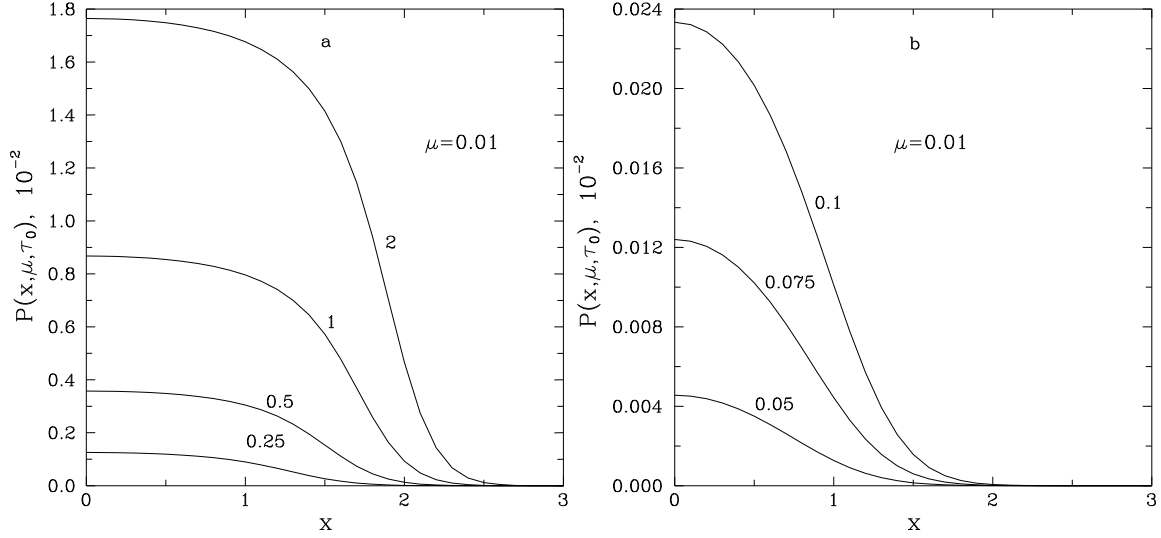}}

\vspace{-11cm}

\caption{Frequency profiles of polarization in a spectral line
for different values of $\tau_0$ marked near the curves.}
\end{figure}

As in the case of Rayleigh scatering we introduce the intensity change and
polarization profiles of the emerging radiation at $\tau=0$ in analogy to eq. 
(\ref{2.12}) where now according to eq. (\ref{3.3})
\begin{equation}
\label{3.7}
\begin{array}{cc}
{\displaystyle
r(x,\mu,\tau_0)=\mu\left[1-e^{-\phi(x)\tau_0/\mu}\right]+
\phi(x)\int_0^{\tau_0}e^{-\phi(x)\tau/\mu}\left[s_{\rm I}(\tau)+
\sqrt{W/8}\,(1-3\mu^2)s_{\rm Q}(\tau)\right]d\tau/\mu,}\\
{\displaystyle
P(x,\mu,\tau_0)=-3\sqrt{W/8}\,(1-\mu^2)\phi(x)\int_0^{\tau_0}e^
{-\phi(x)\tau/\mu}s_{\rm Q}(\tau)d\tau/\mu{.}}\\
\end{array}
\end{equation}

\begin{figure}[t]
\vspace*{-6cm}

\centering
\resizebox{1.0\textwidth}{!}{\includegraphics{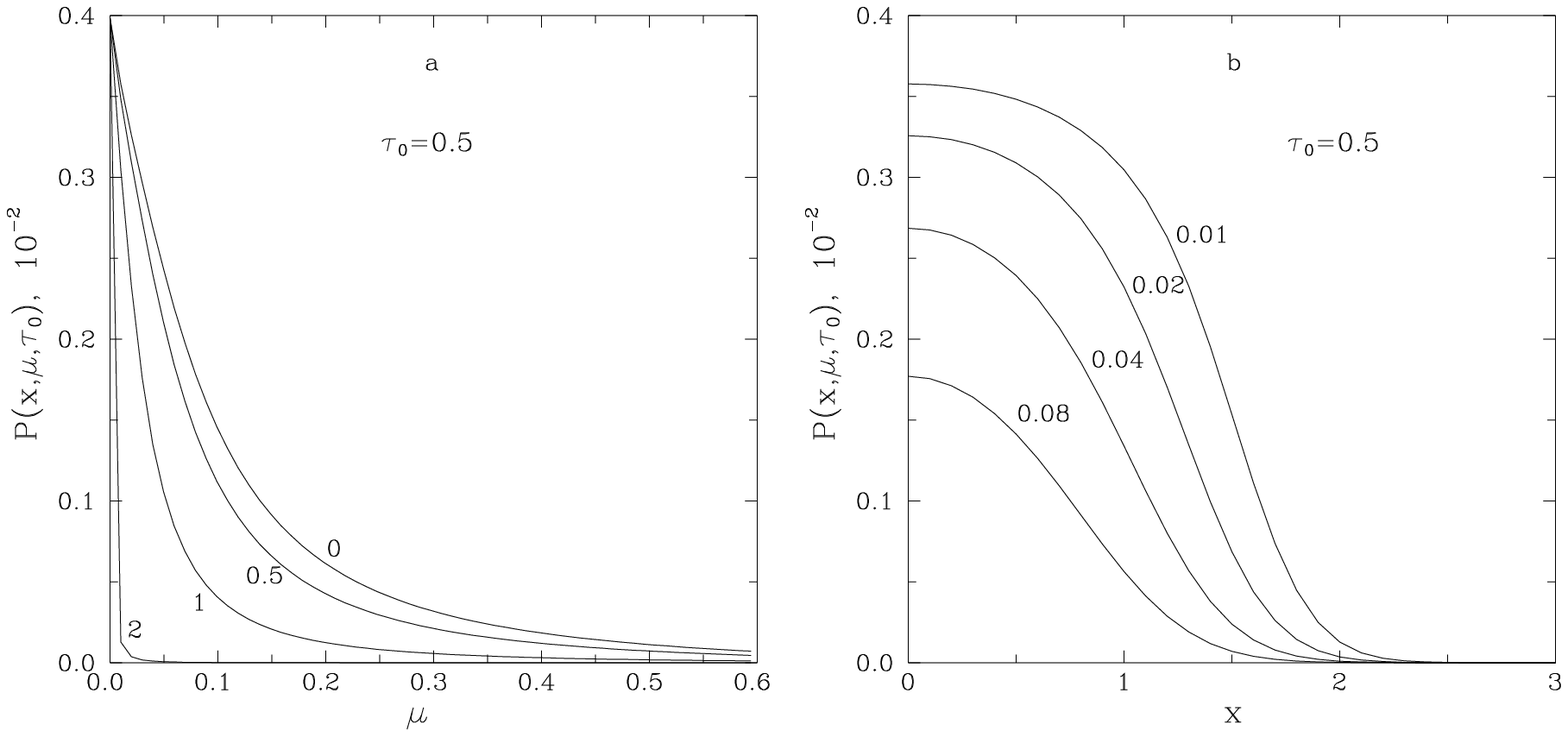}}

\vspace{-11cm}

\caption{Profiles of polarization in a spectral line: a)
 for different 
frequencies $x$ (marked near the curves), b) for different values of $\mu$
(marked near the curves).}
\end{figure}

For the maximum polarization which is reached at $\mu=0$ and for the 
corresponding intensity change we have from above equations
\begin{equation}
\label{3.8}
\begin{array}{cc}
P(x,0,\tau_0)=-3\sqrt{\frac{W}{8}}s_{\rm Q}(0),\\
r(x,0,\tau_0)=s_{\rm I}(0)+\sqrt{\frac{W}{8}}s_{\rm Q}(0)\mbox{.}
\end{array}
\end{equation}
Note that these quantities do not depend on $x$.

The integral equation (\ref{2.9}) with the core matrix and free
term, given by eqs. (\ref{3.6}) and (\ref{3.4}-\ref{3.5}) respectively,
was solved numerically as in the case of Rayleigh scattering. We have used
the Doppler profile $\phi(x)=\pi^{-1/2}e^{-x^2}$ and brought $W=1$, 
$\lambda=1$. These values of parameters $W$ and $\lambda$ evidently maximize 
the degree of polarization.
Fig. 3 shows the dependency of polarization $P$ on the angular 
variable $\mu$ in the
center of the line ($x=0$) for different values of the optical thickness 
$\tau_0$. Comparing Figs.
2 and 3 we see that the resonance polarization is lower a bit than the 
Thomson one,
that is quite expectable. Fig. 4 shows the dependency of $P$ on 
dimensionless
frequency $x$ inside the line for different values of $\tau_0$. We can see 
that with increasing of
$\tau_0$ the line becomes wider, which is natural since the number of 
scatterings rises. The same is
true for intensity. The dependency of polarization on $\mu$ for $\tau_0=0.5$ 
and different
frequencies is shown in the Fig. 5a. We can see that this dependency is very 
sharp in the
wings of the line, which is explained by small optical thickness there. 
Finally, the dependence of
$P$ on $x$ for different values of $\mu$ is shown in the Fig. 5b. With the 
decrease of $\mu$ the line becomes wider since the optical path along the
line of sight increases.

For $\tau_0\ll 1$ we can obtain the following expansions for the components 
of the vector source function:
\begin{equation}
\label{3.9}
\begin{array}{c}
{\displaystyle
s_{\rm I}(\tau)\sim \frac{\lambda}{2}\frac{1}{\sqrt{2\pi}}(\tau_0-2\tau),}\\
{\displaystyle
s_{\rm Q}(\tau)\sim \frac{\lambda}{4\pi}\sqrt{\frac{W}{8}}\left\{\frac{1}
{\sqrt{3}}\left[\tau_0(\tau_0-2\tau)(C-1/6)-
\tau^2\ln\frac{\tau}{\sqrt{\pi}}+(\tau_0-\tau)^2\ln\frac{\tau_0-\tau}{\sqrt{
\pi}}\right]+\right.}\\
{\displaystyle
\left.+\frac{\lambda}{2}\left[\tau(\tau_0-\tau)\ln\frac{\tau_0-\tau}{\tau}+
\frac{\tau_0}{2}(\tau_0-2\tau)\right]\right\}.}\\
\end{array}
\end{equation}

Substituting these expansions into eqs. (\ref{3.7}) and 
(\ref{3.8}) we obtain 
the following expressions for the 
CMB intensity change and polarization 
profiles (for the observer
 reference frame):
\begin{equation}
\label{3.10}
r(x,\mu,\tau_0)\sim\left\{\begin{array}{ll}
(\lambda/2)\tau_0/\sqrt{2\pi},\quad\mu\ll\tau_0\phi(x),\\
\phi(x)\tau_0,\quad\mu\gg\tau_0\phi(x),\\
\end{array}\right.
\end{equation}

\begin{equation}
\label{3.11}
P(x,\mu,\tau_0)\sim\left\{\begin{array}{ll}
{\displaystyle
-\frac{3}{32}\frac{\lambda W}{\pi}\tau_0^2\left[\frac{1}{\sqrt{3}}\left(C-
\frac{1}{6}+
\ln\frac{\tau_0}{\sqrt{\pi}}\right)+\frac{\lambda}{4}\right],\quad\mu\ll
\tau_0\phi(x),}\\
{\displaystyle
-\frac{1}{64}\frac{\lambda W}{\pi}\phi^2(x)\frac{1-\mu^2}{\mu^2}\tau_0^4
\left[\frac{1}{\sqrt{3}}
\left(C-\frac{1}{4}+\ln\frac{\tau_0}{\sqrt{\pi}}\right)+\frac{3}{8}\lambda
\right],\quad\mu\gg\tau_0\phi(x).}\\
\end{array}\right.
\end{equation}

At low values of $\tau_0$ the last formula is in a good agreement with the 
results of numerical calculations. For example, the polarization dependency on
$\tau_0$ is quite sharp ($\sim\tau_0^4$) for $\mu\gg\tau_0\phi(x)$, but is 
more slow ($\sim\tau_0^2$) for $\mu\ll \tau_0\phi(x)$. For approximate 
estimation of
polarization the accuracy of eq. (\ref{3.11}) is quite enough. For example, at
$\tau_0=0.01$ and $\mu >0.1$ its error is less than $3.5\%$.

\section{Conclusion}

In this work we investigate the possible observational display of Thomson 
scattering on free electrons and of resonance scattering in a spectral 
line of cosmic microwave background radiation in the flat moving layers from
``dark ages" epoch ($10<z<1000$). The formulae for appearing anisotropy and 
polarization of CMB are obtained and numerical calculations are made. We show
that for characteristic parameters according to recent models of substance
evolution at that period the values of intensity fluctuations and polarization
of cosmic microwave background may be in the following ranges:
$\Delta I/I_0 = 10^{-4}\div 10^{-6}$, $Q/I=10^{-6}\div 10^{-7}$. The effects 
are observable and their discover would be a sufficient 
forward step in investigations of substance evolution in the pre-galactic 
epoch.

We do not consider a transfer of the calculated intensity and polarization
changes through a homogeneous universe from $z\approx 150$ to $z=0$.
The possible distortions caused by such a transfer are defined by an
optical thickness of the universe due to Thomson scattering. According
to the recent measurements in the BOOMERANG and MAXIMA experiments (see
\cite{23}, \cite{24}) the Thomson optical thickness of the universe between 
$z=0$ and $z=1100$
is evaluated to be less than 0.1. As we conserned by an interval between
$z=0$ and $z\approx 150$ the distortions of the calculated fluctuations
will not exceed probably a few percents which is undoubtedly less than
an uncertainty of the protoobject models.
\medskip

\noindent{\bf Acknowledgements}. We are grateful to Yu. Parijskij, V. Ivanov, 
D. Nagirner and A. Starobinsky for the interest and 
useful discussions. This research is supported in part by
Grant \# 00--15--96607 under the program ``Leading Scientific Schools" and
by Grant \# 02--02--16535 from the Russian Foundation for Basic Research.

\end{document}